\documentclass[useAMS,usenatbib,onecolumn]{mnras}
\usepackage{amsmath,bm}
\usepackage{dcolumn}
\usepackage[utf8]{inputenc}
\usepackage{graphics,graphicx}
\usepackage{float}
\usepackage{threeparttable}
\usepackage{url}
\usepackage{epstopdf}
\usepackage{soul}

\usepackage{color}

\newcommand{\p}{^\prime}
\newcommand{\pp}{^{\prime\prime}}

\title[ExoMol line lists -- XXXVIII. SiO$_2$]{ExoMol line lists -- XXXVIII. High-temperature molecular line list of silicon dioxide (SiO$_{\mathbf{2}}$)}

\date{\today}

\author[A. Owens et al.]
{A. Owens,$^{1}$\thanks{The corresponding author: alec.owens.13@ucl.ac.uk} E.K. Conway,$^{1,2}$ J. Tennyson$^1$\thanks{The corresponding author: j.tennyson@ucl.ac.uk} and S. N. Yurchenko$^1$\thanks{The corresponding author: s.yurchenko@ucl.ac.uk}\vspace*{4mm}\\
$^1$ Department of Physics and Astronomy, University College London, Gower Street, WC1E 6BT London, UK\\
$^2$ Atomic and Molecular Physics Division, Center for Astrophysics $|$ Harvard \& Smithsonian, Cambridge, Massachusetts 02138, USA}

\date{Accepted XXXX. Received XXXX; in original form XXXX}

\pagerange{\pageref{firstpage}--\pageref{lastpage}} \pubyear{2020}

\begin{document}

\label{firstpage}

\maketitle

\begin{abstract}
Silicon dioxide (SiO$_2$) is expected to occur in the atmospheres of hot rocky super-Earth exoplanets but a lack of spectroscopic data is hampering its possible detection. Here, we present the first, comprehensive molecular line list for SiO$_2$. The line list, named OYT3,  covers  the wavenumber range 0\,--\,6000~cm$^{-1}$ (wavelengths $\lambda > 1.67$~$\mu$m) and is suitable for temperatures up to $T=3000$~K. Almost 33 billion transitions involving 5.69 million rotation-vibration states with rotational excitation up to $J=255$ have been computed using robust first-principles methodologies. The OYT3 line list is available from the ExoMol database at \href{http://www.exomol.com}{www.exomol.com}.
\end{abstract}


\section{Introduction}

In the gas phase, silicon dioxide ($^{28}$Si$^{16}$O$_2$) is a linear triatomic 
molecule, analogous to CO$_2$. SiO$_2$ 
is expected to be present in the atmospheres of hot rocky super-Earth 
exoplanets~\citep{jt693}. These tidally-locked exoplanets are in close proximity 
to their host star with their dayside exposed to temperatures reaching 4000~K. At 
such high temperatures the material on the surface of the planet vaporises to 
produce an atmosphere strongly dependent on initial planetary 
composition~\citep{09ScFexx,11MiKaFe}, e.g.\ composed of SiO$_2$-rich silicates 
like the Earth's continental crust~\citep{12ScLoFe.exo}. Furthermore, the likely 
presence of water vapour creates a steam atmosphere, and since all major 
rock-forming elements (Si, Mg, Ca, etc.) dissolve in steam to some extent, one 
can expect to encounter simple molecules composed of rock-forming elements with 
oxygen and hydrogen~\citep{16FeJaWi}.

Investigating the spectroscopy of hot rocky super-Earths requires accurate 
molecular opacities on systems such as SiO$_2$. There is, however, very limited 
information on some of these molecules, partly because they form in the gas 
phase at very high temperatures making their spectra challenging to measure in the laboratory. 
Instead, theory offers a more viable route for generating the molecular line lists of these systems through systematic approaches based on first-principles 
methodologies~\citep{jt654,jt626}. These computational procedures have been 
successfully adopted by the ExoMol database~\citep{jt528,jt631}, which provides 
comprehensive line lists suitable for modelling exoplanet atmospheres at 
elevated temperatures. Already a large number of important diatomic and 
polyatomic species have been treated within the ExoMol framework~\citep{jt731}, 
and efforts are now being focused on molecules relevant to hot rocky super-Earth 
atmospheres. This brings about its own unique set of challenges, notably the 
completeness of the line list at very high temperatures and the lack of 
experimental data to refine the theoretical spectroscopic model.

Regarding SiO$_2$, only a few studies have investigated its infrared spectrum 
with measurements of the $\nu_2$ bending mode~\citep{92AnMcxx.SiO2} and $\nu_3$ 
stretching mode~\citep{78Scxxxx.SiO2,80Scxxxx.SiO2}. However, since these 
studies were performed in solid argon matrices the measured wavenumbers can be 
shifted by tens of wavenumbers~\citep{94Jacox} making it difficult to assess the usefulness of the determined 
values for gas phase studies. Similarly, only a small number of theoretical studies have considered silicon 
dioxide~\citep{67KaMuKl.SiO2,78PaHexx.SiO2,96WaWuDe.SiO2,
99BrTsSc.SiO2,09KoAhMe.SiO2,14HaXiSc.SiO2}, but none of these are particularly 
relevant to the work presented here, namely the high-accuracy calculation of its 
rotation-vibration spectrum.

In this work, we present the first, comprehensive rotation-vibration line list 
of gas-phase SiO$_2$. The new line list has been computed using robust 
first-principles methodologies~\citep{jt654} within the ExoMol computational 
framework~\citep{jt626} and adds to the other available silicon-bearing 
molecules in the ExoMol database: SiH$_4$~\citep{jt701}, SiH~\citep{jt776}, 
SiO~\citep{jt563}, SiS~\citep{jt724} and SiH$_2$~\citep{jt779}.

\section{Methods}
\label{sec:methods}

The computational approach used to produce the SiO$_2$ line list is described in 
detail in the supplementary material and only a brief summary is provided here. 
Initially, high-level \textit{ab initio} methods were used to compute the 
potential energy surface (PES) and dipole moment surface (DMS) of the electronic 
ground state of SiO$_2$. The PES was generated using explicitly correlated 
coupled cluster (CCSD(T)-F12b) calculations with extrapolation to the complete basis set limit, 
and included several additive energy corrections to account for small effects 
like scalar relativity. This approach is capable of producing accurate PESs for 
closed-shell molecules that can reproduce fundamental term values to within $\pm 
1$~cm$^{-1}$ on average (e.g.\ see \citet{19OwYaKu.CH3F} and references within). 
Due to the lack of reliable experimental data for this molecule, the PES was 
not empirically refined. The DMS was computed using CCSD(T)-F12b with a large augmented correlation consistent basis set. 
It is now well established that transition intensities computed using \textit{ab 
initio} DMSs are comparable to, and occasionally more reliable, than 
experiment~\citep{13Yurchenko.method,jt573}. Both surfaces were computed on the 
same grid of 15\,365 nuclear geometries and then fitted with suitable analytic 
representations for use in the next stage of the calculation process. The 
potential energy and dipole moment surfaces are provided as supplementary 
material along with Fortran routines to construct them.

Line list calculations employed the variational nuclear motion program 
\textsc{TROVE}~\citep{TROVE}, which was extended to treat linear triatomic 
molecules in this work. Benchmarking was performed against the triatomic nuclear 
motion code \textsc{DVR3D}~\citep{jt338} to ensure the validity of the 
\textsc{TROVE} implementation. The ability to utilise two nuclear motion codes 
based on different methodologies proved highly beneficial and meant the 
theoretical spectroscopic model of SiO$_2$ could be checked for consistency. 
This was particularly important given the lack of experimental data to compare 
against. A large symmetry-adapted basis set was used in the rovibrational 
calculations with convergence testing performed at different $J$ values.

The line list was computed with a lower state energy threshold of $h c \cdot 
15\,000$~cm$^{-1}$ ($h$ is the Planck constant and $c$ is the speed of light) 
and considered transitions up to $J=255$ in the 0\,--\,6000~cm$^{-1}$ range. The 
energy levels and wavefunctions of SiO$_2$ can be classified under the 
$\bm{C}_{\mathrm{2v}}\mathrm{(M)}$ molecular symmetry group~\citep{98BuJexx}. 
The nuclear spin statistical weights are $g_{\mathrm{ns}}=\lbrace 
1,1,0,0\rbrace$ for states of symmetry $\lbrace A_1,A_2,B_1,B_2\rbrace$, 
respectively.
Thus, $B_1$ and $B_2$ states need not be computed and transitions follow the 
symmetry selection rules $A_1 \leftrightarrow A_2$; and the standard rotational 
selection rules, $J\p-J\pp=0,\pm 1,\; J\p+J\pp \ne 0$; where $\p$ and $\pp$ 
denote the upper and lower state, respectively. These representations are 
correlated to the $\bm{D}_{\mathrm{\infty h}}\mathrm{(M)}$ irreducible 
representation, commonly used for linear molecules, as $A_1 \leftrightarrow 
\Sigma_g^{+}$ and $A_1 \leftrightarrow \Sigma_u^{-}$. Another standard 
spectroscopic  descriptor is the Kronig parity $e/f$ \citep{75BrHoHu.diatom}, related to the total 
parity $+1$ ($A_1$) and  $-1$ ($A_2$) as follows: the parity of the $e$ state is 
$(-1)^J$ while the parity of the $f$ state is $(-1)^{J+1}$.

The vibrational quantum numbers used by TROVE (see supplementary material) were 
correlated to the following standard spectroscopic quantum numbers used for 
linear-type triatomic molecules: $v_1$, $v_2^{\rm lin}$, $L=|l|$, $v_3$, where 
$v_1$ and $v_3$ are the symmetric and asymmetric stretching quantum numbers, respectively, $v_2$ is the bending vibrational quantum number used for linear molecules and  
$l$ is the corresponding vibrational quantum number. The two bending quantum 
numbers $v_2^{\rm lin}$ and $l$ are connected to the `non-linear' bending 
quantum number $v_2$ by $v_2^{\rm lin} = 2 v_2 + l$ with $L = v_2^{\rm lin},  
v_2^{\rm lin} - 2, \ldots 0 (1)$.

The symmetries of the vibrational and rotational contributions span the $A_1$, 
$A_2$, $B_1$ and $B_2$ irreducible representations (irreps) in 
$\bm{C}_{\mathrm{2v}}\mathrm{(M)}$ and $\Sigma_{g/u}^{+/-}$ ($L=0$), $\Pi_{g/u}$ 
($L=1$), $\Delta_{g/u}$ ($L=2$) etc. in $\bm{D}_{\mathrm{\infty h}}$. Odd values 
of the quantum number $v_3$ indicates vibrational states of $B_2$ symmetry. The 
rotational quantum number $k_a$ is constrained to the vibrational angular 
momentum by $k_a=l$.

\begin{figure}
\centering
\includegraphics[width=0.48\textwidth]{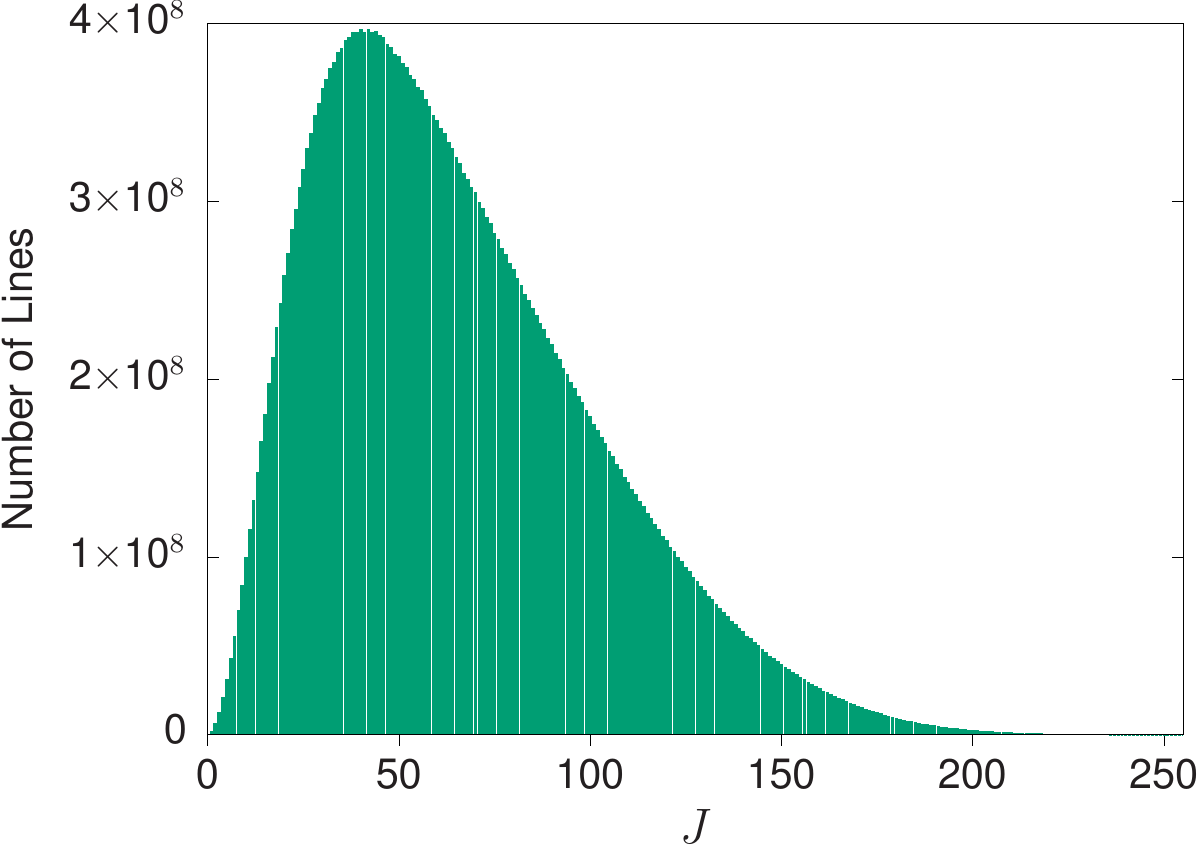}
\includegraphics[width=0.48\textwidth]{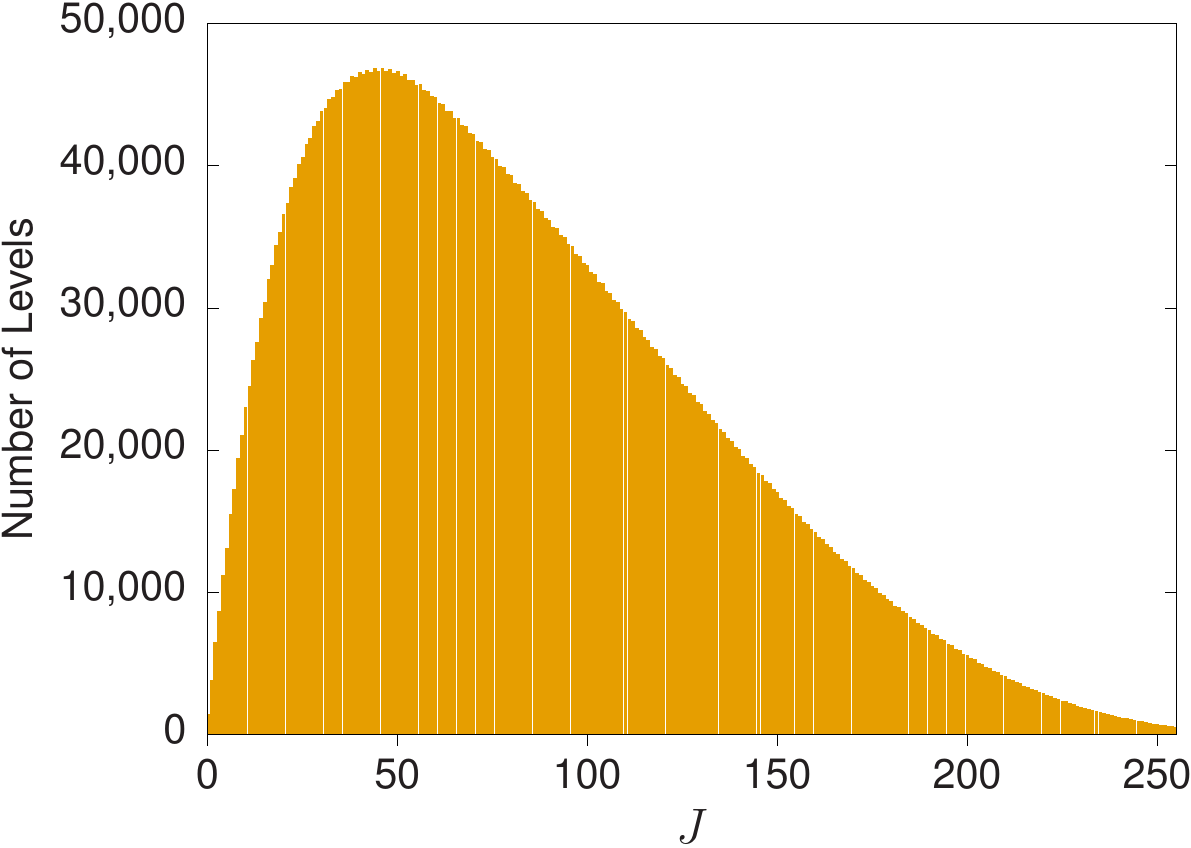}
\caption{\label{fig:lines_levels}The total number of lines (left panel) and energy levels (right panel) for each value of the rotational quantum number $J$ in the OYT3 line list. For the number of lines, a single $J$ value counts transitions between $J\leftrightarrow J\!-\!1$ and $J\leftrightarrow J$.}
\end{figure}

A total of 32\,951\,275\,437 transitions involving 5\,688\,942 energy levels up to $J=255$ were computed for the OYT3 line list. The distribution of lines and levels as a function of $J$ is illustrated in Fig.~\ref{fig:lines_levels}. The largest number of transitions in the OYT3 line list occurs between $J=39\leftrightarrow 40$, while the number of states peaks at $J=46$ before smoothly decreasing, a result of the upper state energy threshold of $h c \cdot 21$\,000\,cm$^{-1}$.

\section{Results}
\label{sec:results}

\subsection{Partition function of silicon dioxide}
\label{sec:pfn}

\begin{figure}
\centering
\includegraphics[width=0.5\textwidth]{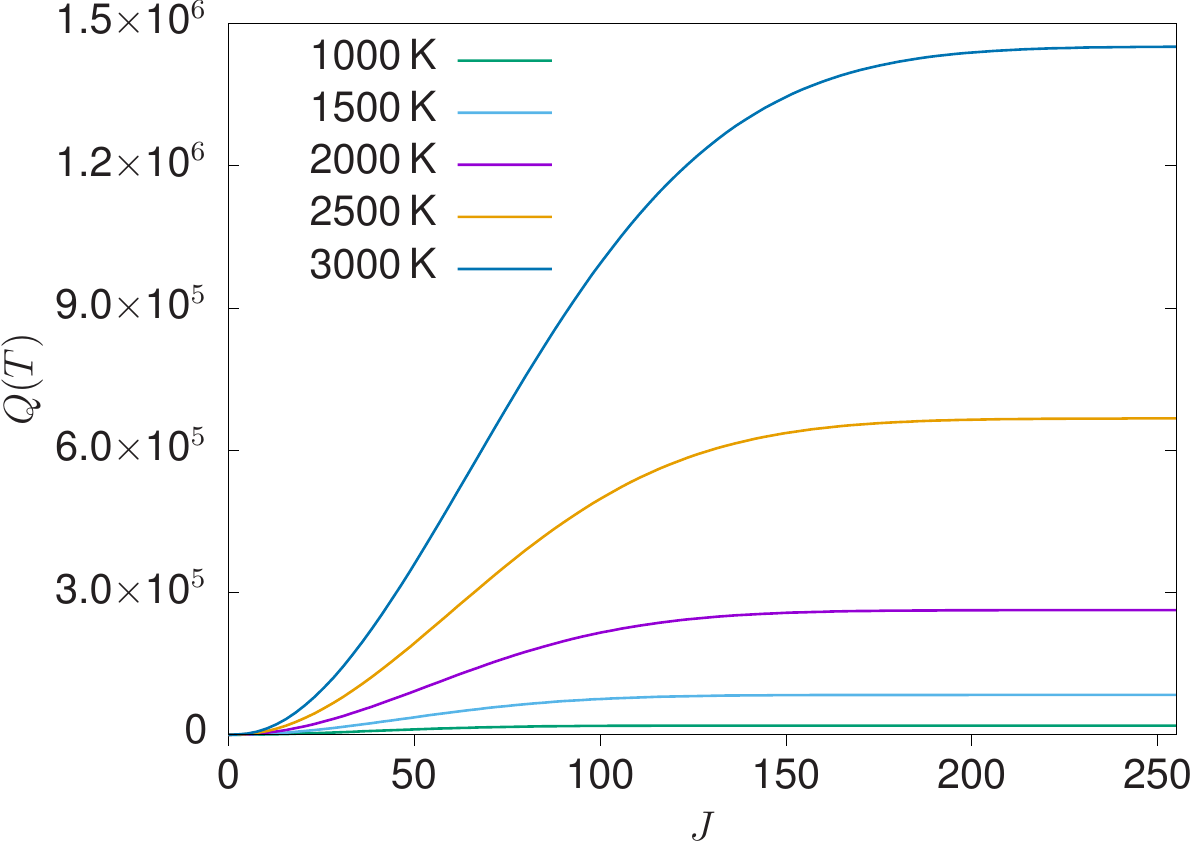}
\caption{\label{fig:pfn}Convergence of the partition function $Q(T)$ of SiO$_2$ with respect to the rotational quantum number $J$ at different temperatures.}
\end{figure}

The temperature-dependent partition function $Q(T)$ is expressed as,
\begin{equation}
\label{eq:pfn}
Q(T)=\sum_{i} g_i \exp\left(\frac{-E_i}{kT}\right) ,
\end{equation}
where $g_i=g_{\rm ns}(2J_i+1)$ is the degeneracy of a state $i$ with energy $E_i$ and rotational quantum number $J_i$. Values of the partition function of SiO$_2$ have been computed by summing over all calculated rovibrational energy levels on a $1$~K grid in the 1\,--\,3000~K range (provided as supplementary material). In Fig.~\ref{fig:pfn}, the convergence of $Q(T)$ as a function of $J$ for select temperatures is shown. At lower temperatures the partition function converges quickly but a substantial number of high $J$ states must be considered to achieve convergence above 1500~K. At $J=255$, the value of $Q(2000\,{\rm K})$ is converged to 0.0001\%, while the value of $Q(3000\,{\rm K})$ is converged to 0.0019\%.

\begin{figure}
\centering
\includegraphics[width=0.5\textwidth]{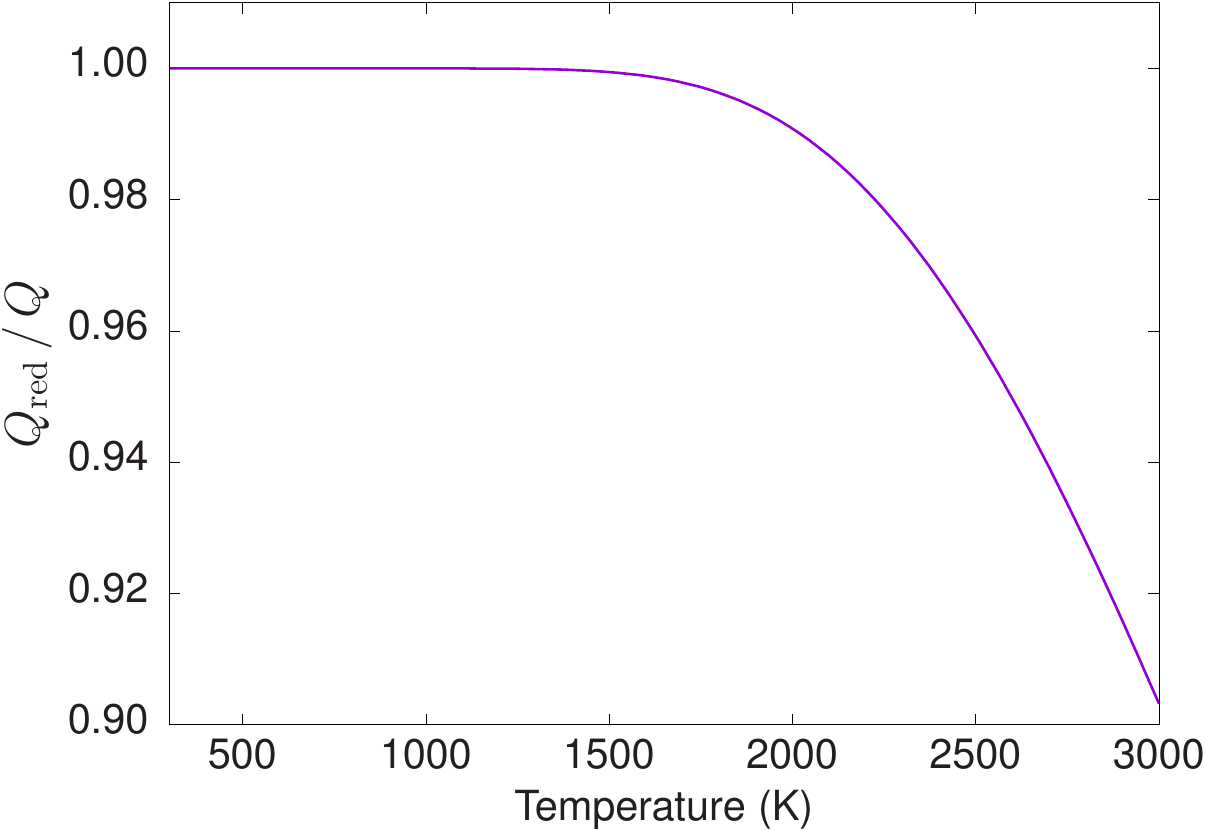}
\caption{\label{fig:pfn_lim}The ratio $Q_{\rm red}/Q$ as a function of temperature $T$; this provides a measure of the completeness of the OYT3 line list.}
\end{figure}

The SiO$_2$ line list was computed with a lower state energy threshold of $h c \cdot 15\,000$~cm$^{-1}$. A measure of the completeness of the line list can be obtained by studying the reduced partition function $Q_{\rm red}(T)$, which only includes energy levels up to $h c \cdot 15\,000$~cm$^{-1}$ in the summation of Eq.~\eqref{eq:pfn}. The ratio $Q_{\rm red}(T)/Q(T)$ has been plotted with respect to temperature in Fig.~\ref{fig:pfn_lim} and from this we see that above 1500~K the ratio starts to decrease from unity. At 3000~K, the ratio $Q_{\rm red}/Q=0.90$ and this should be considered as a soft temperature limit for the OYT3 line list. Uses above this will result in a progressive loss of opacity but the missing opacity contribution can be estimated from $Q_{\rm red}/Q$ if needed~\citep{jt181}.

\subsection{Line list format}

 The SiO$_2$ line list is provided in the ExoMol data format and further details with illustrative examples can be found in \citet{jt631}. The \texttt{.states} file, see Table~\ref{tab:states}, contains all the computed rovibrational energy levels (in cm$^{-1}$), each labelled with a unique state ID counting number, symmetry and quantum number labelling, and the contribution $C_i$ from the largest eigencoefficient used to assign the rovibrational state. The \texttt{.trans} files have been split into $100$~cm$^{-1}$ frequency bins for user-handling purposes and contain all computed transitions with upper and lower state ID labels and Einstein $A$ coefficients, see Table~\ref{tab:trans}.

\begin{table*}
\caption{\label{tab:states}Extract from the \texttt{.states} file of the SiO$_2$ OYT3 line list.}
\begin{threeparttable}
{\tt
\centering
\tabcolsep=5pt
\begin{tabular}{rrcclcccccccccccccc}
\hline\hline\\[-3mm]
\multicolumn{1}{r}{$i$} & \multicolumn{1}{c}{$\tilde{E}$} & $g_{\rm tot}$ & $J$ & unc & $\Gamma_{\rm tot}$& $e/f$ & $v_1$ & $v_2^{\rm lin}$ & $L$ &  $v_3$ &$C_i$ & $n_1$ & $n_2$ & $n_3$ & $\Gamma_{\rm vib}$  & $K$ &  $\Gamma_{\rm rot}$\\
\hline\\[-3mm]
           1 &      0.000000   &   1   &    0 & 0.000000 &A1 & e &   0  & 0 &  0 &  0 &  1.00  &    0 &  0 &  0& A1 &     0& A1\\
           2 &  578.229349   &   1   &    0 & 2.000000&A1 & e &   0  & 2 &  0 &  0 &  1.00  &    0 &  0 &  1& A1 &     0& A1\\
           3 &  990.856966   &   1   &    0 & 2.000000&A1 & e &   1  & 0 &  0 &  0 &  1.00  &    1 &  0 &  0& A1 &     0& A1\\
           4 & 1154.529779   &   1   &    0 & 4.000000&A1 & e &   0  & 4 &  0 &  0 &  1.00  &    0 &  0 &  2& A1 &     0& A1\\
           5 & 1570.947837   &   1   &    0 & 4.000000&A1 & e &   1  & 2 &  0 &  0 &  1.00  &    1 &  0 &  1& A1 &     0& A1\\
           6 & 1728.958462   &   1   &    0 & 6.000000&A1 & e &   0 & 6 &  0 &  0 &  1.00  &    0 &  0 &  3& A1 &     0& A1\\
           7 & 1977.509720   &   1   &    0 & 4.000000&A1 & e &   2  & 0 &  0 &  0 &  1.00  &    1 &  1 &  0& A1 &     0& A1\\
           8 & 2148.937437   &   1   &    0 & 6.000000&A1 & e &   1  & 4 &  0 &  0 &  1.00  &    1 &  0 &  2& A1 &     0& A1\\
           9 & 2301.592712   &   1   &    0 & 8.000000&A1 & e &   0  & 8 &  0 &  0 &  1.00  &    0 &  0 &  4& A1 &     0& A1\\
         10 & 2559.419897   &   1   &    0 & 6.000000&A1 & e &   2  & 2 &  0 &  0 &  1.00  &    1 &  1 &  1& A1 &     0& A1\\
\hline\hline
\end{tabular}
}
\begin{tablenotes}
\item $i$: State counting number;
\item $\tilde{E}$: Term value (in cm$^{-1}$);
\item $g_{\rm tot}$: Total state degeneracy;
\item $J$: Total angular momentum quantum number;
\item unc: Estimated uncertainty of energy level (in cm$^{-1}$);
\item $\Gamma_{\rm tot}$: Overall symmetry in $\bm{C}_{\mathrm{2v}}\mathrm{(M)}$ ($A_1$ or $A_2$);
\item $e/f$: The Kronig (rotationless) parity;
\item $v_1$, $v_2^{\rm lin}$, $L$, $v_3$: Linear-molecule vibrational quantum numbers;
\item $C_i$: Largest coefficient used in the \textsc{TROVE} assignment;
\item $n_1$\,--\,$n_3$: \textsc{TROVE} vibrational quantum numbers;
\item $\Gamma_{\rm vib}$: Symmetry of the vibrational contribution in $\bm{C}_{\mathrm{2v}}\mathrm{(M)}$;
\item $K$: Rotational quantum number, projection of $J$ onto molecule-fixed $z$ axis ($K=L$);
\item $\Gamma_{\rm rot}$: Symmetry of the rotational contribution in $\bm{C}_{\mathrm{2v}}\mathrm{(M)}$.
\end{tablenotes}
\end{threeparttable}
\end{table*}

\begin{table}
\centering
{\tt
\tabcolsep=10pt
\caption{\label{tab:trans} Extract from the \texttt{.trans} file for the $0$\,--\,$100\,$cm$^{-1}$ window of the SiO$_2$ OYT3 line list.}
\begin{tabular}{rrr}
\hline\hline\\[-3mm]
\multicolumn{1}{c}{$f$}  &  \multicolumn{1}{c}{$i$} & \multicolumn{1}{c}{$A_{if}$}\\
\hline\\[-3mm]
     1901572   &   1832882 & 8.4692e-44\\
     2283835  &   2261046 & 9.7538e-45\\
      948596   &   1016760 & 9.4147e-35\\
     1830303   &   1853859 & 2.7761e-42\\
     4120223    &  4135284 & 1.1104e-41\\
      649389    &   670531 & 1.0237e-43\\
      284492    &   332334 & 2.6832e-31\\
     3288298   &   3306461 & 1.3021e-31\\
     3796784   &   3812806 & 2.0901e-39\\
      366209    &   348742 & 3.7795e-33\\
\hline\hline\\[-2mm]
\end{tabular}
}
\\
\noindent
\footnotesize{
$f$: Upper  state ID; $i$:  Lower state ID; \\
$A_{if}$:  Einstein $A$ coefficient (in s$^{-1}$).
}
\end{table}

The assignment of the vibrational quantum numbers to each state is performed in \textsc{TROVE} by analysing the contribution from the primitive basis functions of the different modes, which  are then converted to the linear-molecule, normal mode quantum numbers (see Table~\ref{tab:states}).
 The connection between the assignment and the primitive basis functions is not always straightforward due to the complicated contraction scheme used to build the final symmetrized rovibrational basis set~\citep{17YuYaOv.methods}. Thus, in instances where the eigen-coefficient $|C_i|$ is very small the assignment should be considered as indicative.
 
The computed energy levels in the OYT3 line list have been assigned uncertainties (in cm$^{-1}$) in the following way: The three fundamental term values have been given an estimated uncertainty of 2~cm$^{-1}$, which has then been propagated to all overtone and combination bands using the TROVE normal mode quantum numbers. For example, a state with ($n_1=5$, $n_2=2$, $n_3=1$) has an estimated uncertainty of 16~cm$^{-1}$. The initial uncertainty estimate of 2~cm$^{-1}$ for the fundamentals is based on our previous experience using similar levels of \textit{ab initio} theory to construct closed-shell molecule potential energy surfaces~\citep{jt612,15OwYuYa.SiH4,jt652,19OwYaKu.CH3F,19OwYuxx.P2H2}. This uncertainty scheme is approximate and should not be relied on in instances where the eigen-coefficient $|C_i|$ is small. We have also opted to round the estimated uncertainties to integer values so that they can be easily differentiated from more robust uncertainties, e.g.\ derived from experiment. Note that the ground state $J>0$ rovibrational term values, i.e.\ ($n_1=0$, $n_2=0$, $n_3=0$) have been assigned uncertainties of 2~cm$^{-1}$  throughout.

\subsection{Simulated spectra of silicon dioxide}
\label{sec:linelist}

\begin{figure}
\centering
\includegraphics[width=0.7\textwidth]{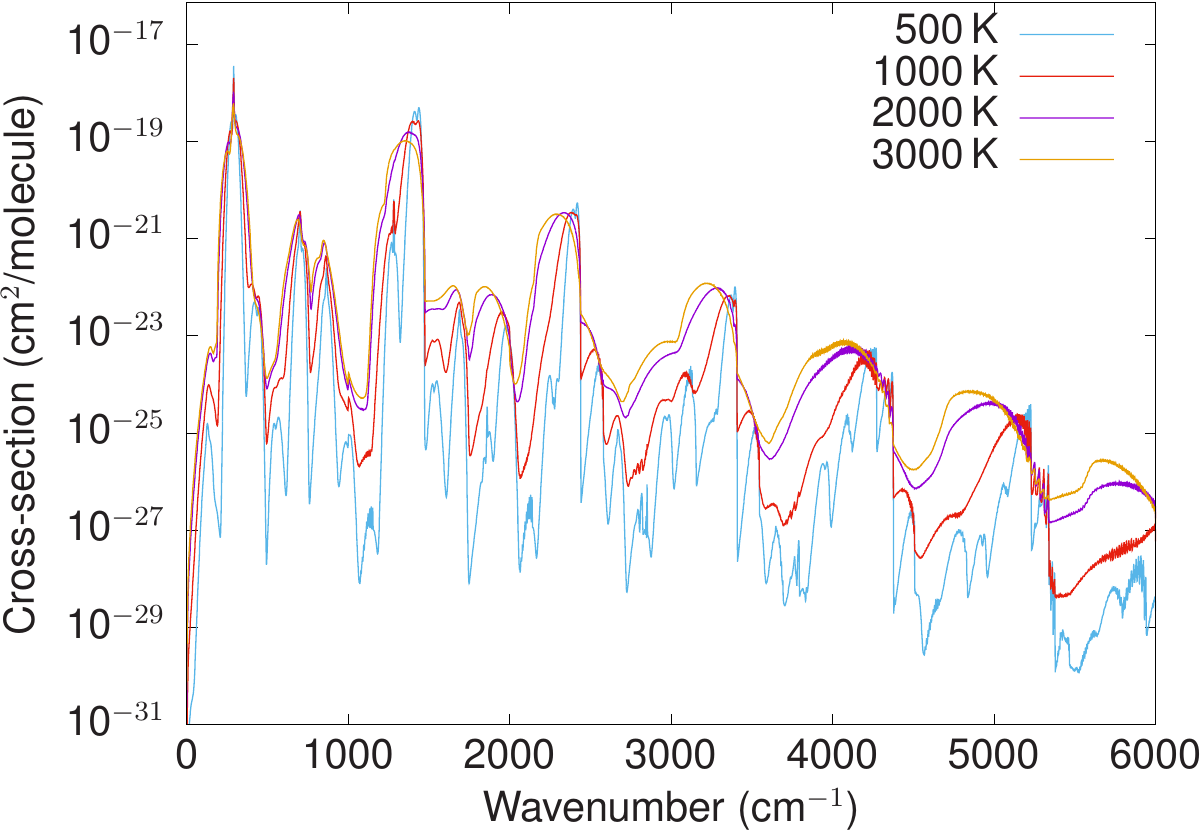}
\caption{\label{fig:temp_sio2}Temperature dependence of the spectrum of SiO$_2$, which becomes increasingly flat as the temperature increases. Absorption cross-sections were computed from the OYT3 line list and represented with a Gaussian line profile with a half width at half maximum (HWHM) of 1~cm$^{-1}$ at a resolution of 1~cm$^{-1}$.}
\end{figure}

The temperature dependence of the OYT3 line list is illustrated in Fig.~\ref{fig:temp_sio2}, where we have plotted integrated absorption cross-sections at a resolution of $1$~cm$^{-1}$ using a Gaussian profile with a half width at half maximum (HWHM) of $1$~cm$^{-1}$. Spectral simulations were performed with the \textsc{ExoCross} program~\citep{jt708}. As expected, the SiO$_2$ spectrum becomes smoother and more featureless as the temperature increases. This is caused by the increased population of vibrationally excited states with temperature, leading to substantial broadening of the rotational band envelopes.

\begin{figure}
\centering
\includegraphics[width=0.49\textwidth]{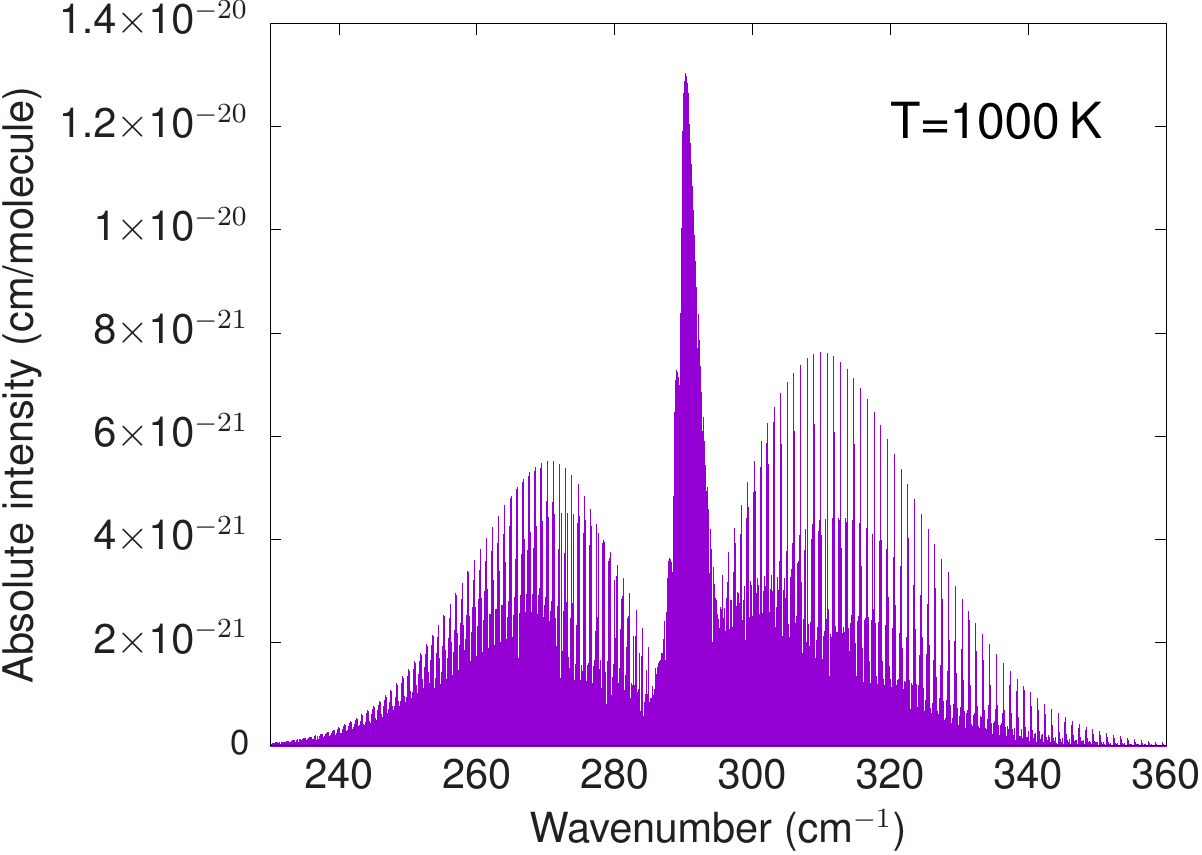}
\includegraphics[width=0.49\textwidth]{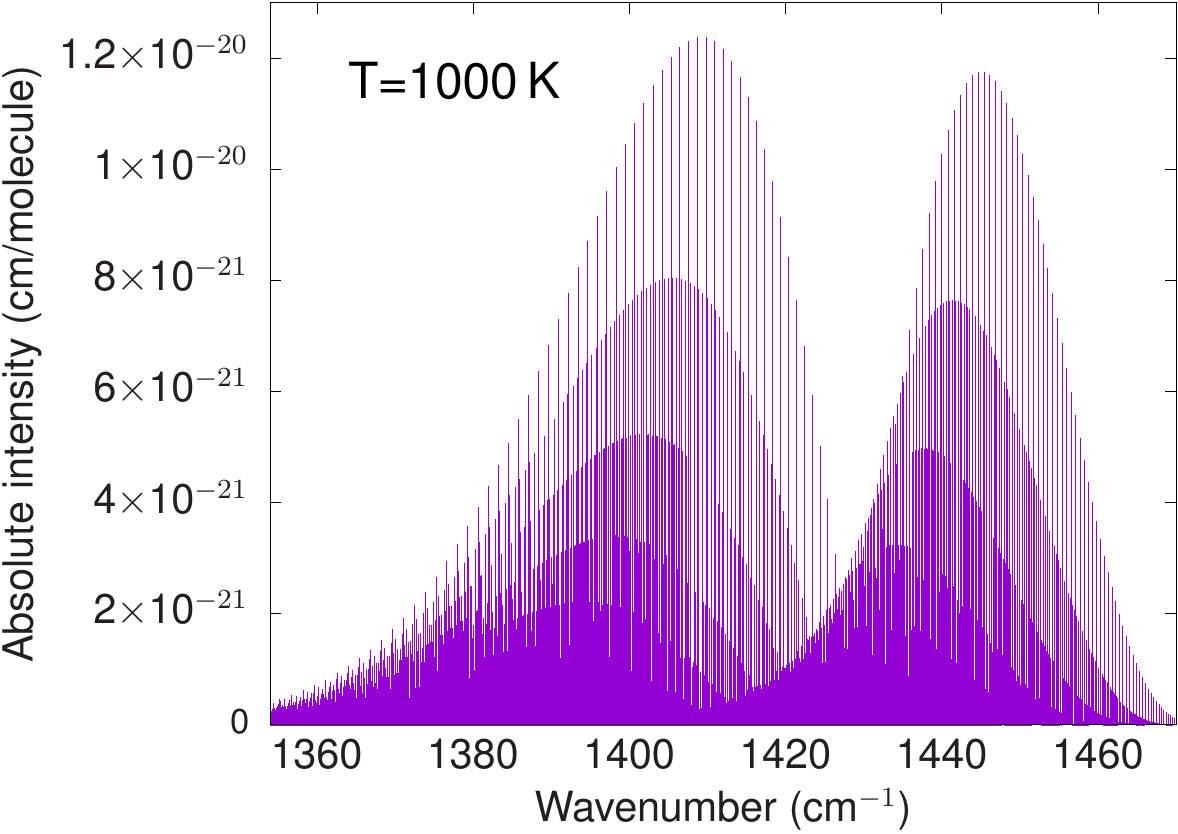}
\caption{\label{fig:stick_sio2}Stick spectrum of the two strongest bands of SiO$_2$ at $T=1000$~K. The left panel shows the bending $\nu_2$ (0,1,1,0) fundamental band while the right panel shows the stretching $\nu_3$ (0,0,0,1) band, where states are labelled by the quantum numbers ($v_1,v_2,L,v_3$).}
\end{figure}

Stick spectra of the two strongest bands at a temperature of 1000~K are shown in Fig.~\ref{fig:stick_sio2}, with the fundamental $\nu_2$ bending mode (left-hand panel) and the stretching $\nu_3$ band (right-hand panel) displayed. The position of these bands is in broad agreement with previous experimental studies~\citep{92AnMcxx.SiO2,78Scxxxx.SiO2,80Scxxxx.SiO2} but given these infrared measurements were performed in solid argon matrices, which are known to shift the measured wavenumbers, we avoid a direct comparison. 

\begin{figure}
\centering
\includegraphics[width=0.95\textwidth]{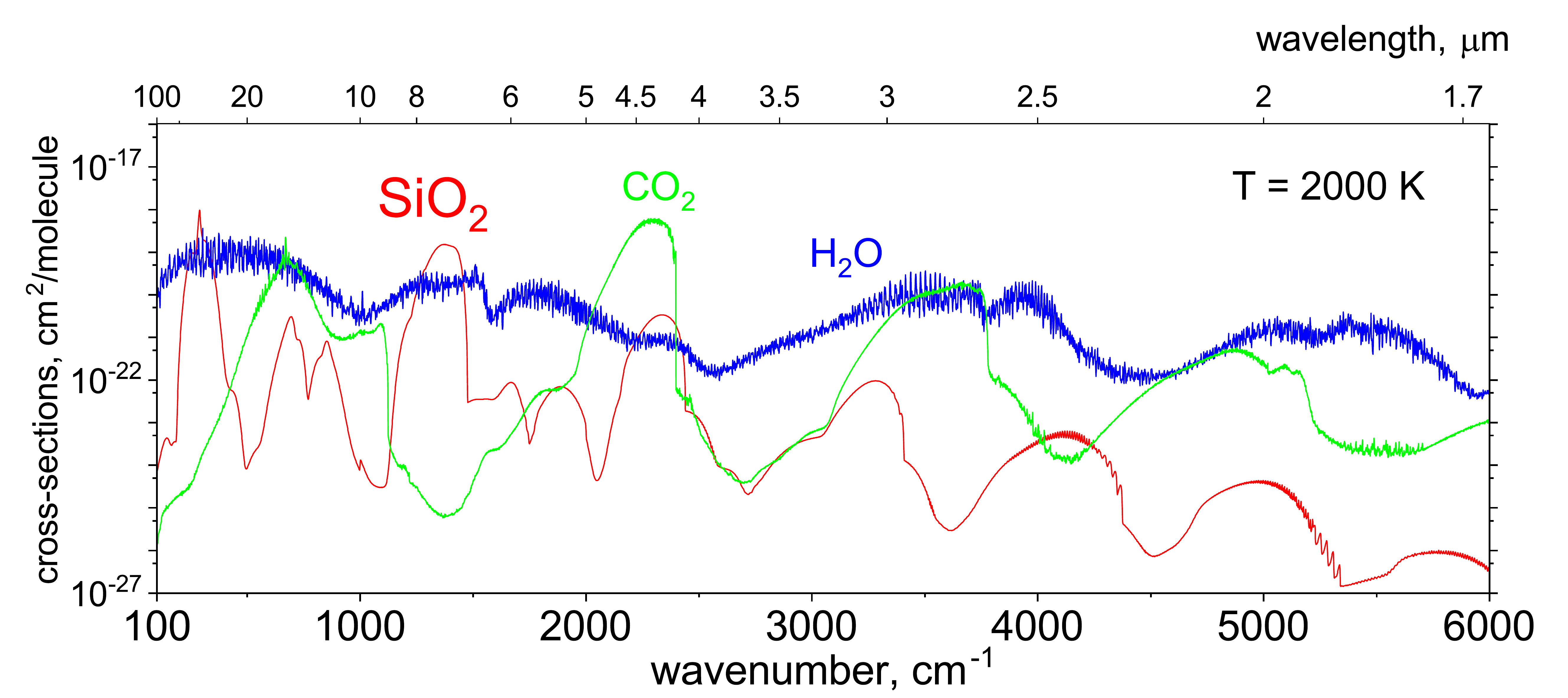}
\caption{\label{fig:CO2-SiO2}Comparison of the 2000~K broadband spectra of SiO$_2$, CO$_2$ and water. A Gaussian line profile with HWHM of 1~cm$^{-1}$ at a resolution of 1~cm$^{-1}$ was used. The CO$_2$ cross sections were computed using the new UCL-4000 line list \protect{\citep{jt805}} and Pokazatel line list for water \protect{\citep{jt734}}. For clarity the relatively flat water spectrum has been reduced by a factor of ten. SiO$_2$ shows strong absorption features around 250 and 1500 cm$^{-1}$. A weaker SiO$_2$  feature at about 2300 cm$^{-1}$ is masked by strong CO$_2$ absorption in this region.}
\end{figure}

The SiO$_2$ spectrum has a distinct strong feature at 4.5~$\mu$m. This region 
plays an important role in the atmospheric applications of exoplanets due to the 
CO$_2$ photometric band used by the Spitzer Space Telescope (IRAC instrument). 
For example, this band was used to build the phase curve of super-Earth 
55~Cancri~e by \citet{16DeGide} and to provide analysis of the atmospheric 
(day/night) structure of the plane. The atmosphere was shown to have a high 
temperature contrast, from  1400~K (night side) to 2700~K (day side). In 
Fig.~\ref{fig:CO2-SiO2}, we show that the CO$_2$ 4.5~$\mu$m region has strong overlap 
with the spectrum of SiO$_2$, namely the $(1,4^0,0)$ band in the same region, 
although the strongest absorption bands are $(0,1^1,0)$ and $(0,2^2,0)$ where 
states are labelled by the quantum numbers ($v_1,v_2^L,v_3$). An absorption spectrum of water is also shown for comparison.
Silicon dioxide has also been suggested as a potential constituent of the 
atmosphere of the super-Earth Corot-7b~\citep{12ScLoFe.exo} and thus should be 
included in retrievals for hot super-Earths.

\section{Conclusion}
\label{sec:conc}

A comprehensive molecular line list for SiO$_2$ has been presented. The line list, named OYT3, covers the 0\,--\,6000~cm$^{-1}$ range (wavelengths $\lambda > 1.67$~$\mu$m) for states below $J=255$ and is applicable for temperatures up to 3000~K. As discussed above, the lack of reliable experimental spectroscopic data on SiO$_2$ has meant that the OYT3 line list has been constructed using purely \textit{ab initio} methods with no degree of empirical refinement. The accuracy of the predicted line positions will suffer as a result, particularly for highly excited states and shorter wavelengths, but for the fundamental bands, which have the strongest intensity, the errors should be within 1--3~cm$^{-1}$ as a conservative estimate. The computed line intensities should not be overly affected and are largely expected to be within the 5--10\% of experimentally determined intensities. Of course, without reliable experimental data to compare against these are only estimates based on our previous experience constructing \textit{ab initio} spectroscopic models with similar electronic structure methods (see, for example, our work on SiH$_4$~\citep{15OwYuYa.SiH4}.

The usual ExoMol methodology is to take advantage of laboratory measurements to improve the accuracy of our computed line lists \citep{jt511}.
However, in the absence of high-resolution spectroscopic measurements for SiO$_2$, this has not proved possible. In practice a number
of current line lists are entirely {\it ab initio}. Of course, for few electron systems such as HD$^+$ \citep{19AmDiJo}, HD \citep{19AmDiJo}, 
HeH$^+$ \citep{jt347,19AmDiJo}, H$_3^+$ \citep{jt666} and LiH \citep{jt506} it is possible to compute high accuracy line lists which should
reproduce astronomical spectra within their observational accuracy. More pertinent here is the case of HCN. The original HCN / HNC line list
of \citet{jt298} was based on a purely {\it ab initio} potential energy and dipole surfaces  \citep{jt273} computed at a lower level
of theory, and hence of  lower accuracy,
than that employed here. This line list 
has been successively updated \citep{jt374,jt570}, and even adapted to H$^{13}$CN \citep{jt447}, by the {\it post hoc} insertion of empirical energy levels,
something that is explicitly allowed for in the ExoMol data format \citep{jt548}. The HCN line list both in its original and updated forms
has proved to be highly useful and indeed underpins a number of recent (possible) detections of HCN in exoplanets \citep{jt629,18HaMaCa.HCN,jt782} and much exoplanet modelling;  it has also been found useful for combustion studies \citep{17GlMaxx.HCN}.
We would anticipate the OYT3 SiO$_2$ line list being used in a similar fashion and, should high resolution SiO$_2$ spectra  become
available, we will update the line list to improve its accuracy. However, we are unaware of any such studies in progress at present.

For present use, we recommend the OYT3 line list for low-resolution studies of exoplanet atmospheres, e.g.\ with a resolving power of $R \approx 100$, but we emphasise that the line list is not designed for high-resolution analysis. Interestingly, a recent cold molecular beam study has shown that silicon dioxide can be efficiently formed through the reaction of SiH and O$_2$ under single collision conditions~\citep{18YaThDa.SiO2}, demonstrating a low-temperature pathway to gas-phase SiO$_2$ that is plausible in the interstellar medium or molecular clouds. The OYT3 line list will aid other astronomical searches for SiO$_2$ and may find use in industrial processes, for example, in the semiconductor industry where silicon-bearing molecules are commonly encountered.

\section*{Acknowledgments}

This work was supported by the STFC Projects No. ST/M001334/1 and ST/R000476/1. The authors acknowledge the use of the UCL Legion High Performance Computing Facility (Legion@UCL) and associated support services in the completion of this work, along with the Cambridge Service for Data Driven Discovery (CSD3), part of which is operated by the University of Cambridge Research Computing on behalf of the STFC DiRAC HPC Facility (www.dirac.ac.uk). The DiRAC component of CSD3 was funded by BEIS capital funding via STFC capital grants ST/P002307/1 and ST/R002452/1 and STFC operations grant ST/R00689X/1. DiRAC is part of the National e-Infrastructure.

\bibliographystyle{mn2e}
\bibliography{journals_astro,additional,jtj,HCN,SiH4,CH3F,P2H2,H2,methods,abinitio,H2CS,H2S,H2O,NH3,programs,SiO2,Books,CH4,exoplanets,diatomic,linelists,matrix}

\section*{Supporting Information}
Supplementary data are available at MNRAS online. This includes the potential energy and dipole moment surfaces of SiO$_2$ 
with programs to construct them, and values of the temperature-dependent partition function of SiO$_2$ up to 3000~K. 
The following references were cited in the supplementary material: \citep{jt612,15OwYuYa.SiH4,jt652,19OwYaKu.CH3F,19OwYuxx.P2H2,98CsAlSc,07AdKnWe.ai,08PeAdWe.ai,09HiPeKn.ai,04TenNo.ai,08YoPexx.ai,02Weigend.ai,05Hattig.ai,MOLPRO,11YaYuRi.H2CS,10HiMaPe.ai,05KaGaxx.ai,08KaGaxx.ai,mrcc_ao,cfour_ao,89Dunning.ai,74DoKrxx.ai,86Hess.ai,01deHaDi.ai,01TyTaSc.H2S,97PaScxx.H2O,03Watson.methods,93JoJexx.H2O,TROVE,jt466,15YaYuxx.method,jt626,
17YuYaOv.methods,jt730,83CaHaSu,jt96,24Numerov.method,61Cooley.method,jt46,jt338,jt114,jt14,jt708}

\label{lastpage}

\end{document}